\documentclass[preprint]{aastex}
\usepackage{mkfig}

\begin{document}

\title{\bf The Mg~II and Lyman-$\alpha$ Lines of Nearby K Dwarfs:
  ISM Components and Flux Measurements\altaffilmark{1}}

\author{Brian E. Wood,\altaffilmark{2} Carol W. Ambruster,\altaffilmark{3}
  Alexander Brown,\altaffilmark{4} and Jeffrey L. Linsky\altaffilmark{2}}

\altaffiltext{1}{Based on observations with the NASA/ESA Hubble Space
  Telescope, obtained at the Space Telescope Science Institute, which is
  operated by the Association of Universities for Research in Astronomy,
  Inc.\ under NASA contract NAS5-26555.}
\altaffiltext{2}{JILA, University of Colorado and NIST, Boulder, CO
  80309-0440; woodb@marmoset.colorado.edu, jlinsky@jila.colorado.edu.}
\altaffiltext{3}{Department of Astronomy and Astrophysics, Villanova
  University, Villanova, PA 19085; ambruster@ucis.vill.edu.}
\altaffiltext{4}{Center for Astrophysics and Space Astronomy, University of
  Colorado, Boulder, CO 80309-0389; ab@casa.colorado.edu}

\begin{abstract}

      We analyze local ISM absorption observed in the Lyman-$\alpha$
and Mg~II h \& k lines of six nearby K dwarf stars, using UV spectra of these
stars obtained with the Goddard High Resolution Spectrograph
on the {\em Hubble Space Telescope}.  For four of the six
stars, we detect an absorption component with a velocity and column density
consistent with the Local Interstellar Cloud (LIC).  For HD~197890, there is
no observed component at the expected LIC velocity, or at the projected
velocity of the G cloud, which is a nearby cloud in the general direction of
the Galactic Center.  It also seems doubtful that either of the two
components seen toward HD~82558 are LIC or G cloud absorption.  The total
H~I column density toward HD~82558 ($d=18.3$~pc) is extremely high
($\log N_{\rm H}=19.05\pm 0.15$), representing the largest average H~I
density detected for any line of sight through the nearby LISM
($n_{\rm H}\approx 0.2$ cm$^{-3}$).  This is particularly remarkable
considering that this star is only $39^{\circ}$ from the
``interstellar tunnel'' toward $\epsilon$~CMa, where column densities are
an order of magnitude lower than this toward stars which are an order of
magnitude farther away.

\end{abstract}


\section{Introduction}

     In 1995--1996, the Goddard High Resolution Spectrograph (GHRS)
on board the {\em Hubble Space Telescope} (HST) observed a
selection of six K dwarf stars with spectral types between K0~V and K2~V,
all but one of which were also observed by the {\em Extreme Ultraviolet
Explorer} (EUVE).  These data were obtained to study how stellar atmospheric
structure varies with rotation rate, as the only major characteristic
distinguishing these stars is rotation period, which ranges from 0.38 to
6.76 days.  Chromospheric and transition region (TR) plasmas ($T<10^{6}$ K)
produce many emission lines detectable in the UV spectra obtained
by GHRS, while EUVE spectra contain many emission line diagnostics of
coronal plasma ($T\geq 10^{6}$ K).  Thus, the combination of the two data
sets provides a complete picture of how the chromospheric, TR, and coronal
plasma emissions vary with rotation for our sample of K dwarfs.  Some
preliminary results of this study were presented by \citet{cwa98}.

     In order to derive accurate emission line fluxes from the GHRS and
EUVE data, it is necessary to correct for interstellar absorption.  The
EUVE fluxes are most affected by the ISM because of the H~I photoionization
edge at 912~\AA\ and the He~I photoionization edge at 504~\AA, which result
in significant continuum absorption in the EUVE bandpasses for even modest
ISM column densities \citep{tr94}.  In our GHRS data, three of the strongest
and most important chromospheric lines show strong absorption features from
the local interstellar medium (LISM) --- H~I Lyman-$\alpha$ at 1216~\AA, and
the Mg~II h \& k lines at 2803~\AA\ and 2796~\AA, respectively.  In this
paper, we analyze the interstellar absorption observed in the Lyman-$\alpha$
and Mg~II lines of the K dwarfs.  In doing so, we hope to measure accurate
stellar Lyman-$\alpha$ and Mg~II line fluxes after removing the LISM
absorption, and also to provide measurements of H~I column densities that
are crucial for the analysis of the EUVE data.

     However, the LISM measurements are also interesting in their own right.
Recent studies of Lyman-$\alpha$, Mg~II, and other UV absorption lines have
greatly improved our knowledge of the physical properties of the LISM,
including measurements of the cosmologically important local D/H ratio
\citep{wl95,jll95,ard97,np97,jll98,mss99}, the ionization state of the local
cloud \citep{bew97,ebj00}, and detections of heliospheric and astrospheric
material heated by interactions between the LISM and solar/stellar winds
\citep{jll96,bew96b,bew98}.  High quality UV spectra provided by HST have
also been used to construct the first crude three-dimensional models for the
distribution of H~I in the immediate vicinity of the Sun \citep{jll00,sr00}.
The K dwarf spectra, which are all of relatively short lines of sight, will
provide more valuable data points for this continuing project.

\section{Observations}

     The six Pleiades Moving Group stars comprising our primary data set are
listed in Table 1 together with their spectral types, {\em Hipparcos}
distances \citep{macp97}, Galactic coordinates, and rotation periods.
The stars are listed in order of increasing rotation period.
In this paper, we also measure the Lyman-$\alpha$ and Mg~II fluxes of some
other early K dwarfs whose interstellar absorption has been observed by HST
and analyzed in the past, and these stars are also listed in Table 1.
The references for the spectral types and rotation periods are provided in
Table 1.  Note, however, that rotation rates can be slightly variable due to
the effects of differential rotation \citep[e.g.,][]{rad96,shs97}.

     The GHRS observations analyzed in this paper are tabulated in Table 2.
A full description of the GHRS instrument is provided by \citet{jcb94}
and \citet{srh95}.  For each star, the G270M and G160M gratings
are used to take moderate resolution spectra
($\Delta\lambda/\lambda \approx 20,000$) of the Mg~II h \& k and
Lyman-$\alpha$ spectral regions, respectively.  The observing program
includes observations of other spectral regions, but we will focus here only
on the spectra of Lyman-$\alpha$ and Mg~II.

     The spectra listed in Table 2 were taken through the small science
aperture (SSA) to maximize the spectral resolution and, more importantly for
the Lyman-$\alpha$ spectrum, to minimize the amount of geocoronal
Lyman-$\alpha$ emission.  For all of our Lyman-$\alpha$ data, the geocoronal
emission feature (when visible at all) is well within the core of the broad,
saturated LISM absorption line.  This makes it easy to remove,
which we do by fitting a Gaussian to the feature and then subtracting the
Gaussian from the data.  Each spectrum was taken in four sequential readouts
using the FP-SPLIT option, in which the spectrum is dithered on the detector
to allow a better correction for fixed pattern noise \citep{srh95}.
The data were reduced using the IDL-based CALHRS software developed
by the GHRS team \citep{rdr92}.

     The last column of Table 2 indicates whether or not a special wavelength
calibration observation of the Pt calibration lamp was taken along with the
observation to allow for a more accurate wavelength calibration.  When such
an observation is available, we use it to determine a better zero-point
offset correction for the wavelengths of the associated science spectrum.
When such an observation is not available, an offset correction is provided by
the so-called SPYBAL (``spectrum y-balance'') observations of the calibration
lamp.  The SPYBALs are used to properly center spectra on the diode array,
and are taken with the same grating as the accompanying science observation,
but generally of a different spectral region.  \citet{drs93}
describe how the SPYBALs can also be used to provide offset
corrections for the wavelengths of science observations.  Even with these
corrections, we do not expect the uncertainties in the wavelength scales of
our SPYBAL-calibrated spectra to be lower than about $\pm 3$ km~s$^{-1}$.

\section{Analyzing the Interstellar Absorption}

\subsection{Modelling the Mg~II Line Profiles}

     Figure 1 shows the Mg~II h \& k lines observed from the K dwarfs.  The
rest wavelengths in air of the h \& k lines are 2802.705~\AA\ and 2795.528~\AA,
respectively.  Interstellar absorption is clearly seen contaminating both
lines, with the stronger absorption in the k line due to a larger oscillator
absorption strength for that line.  In Figure 2, the Mg~II h lines are
displayed on a heliocentric velocity scale, along with our best fits to the
interstellar absorption, which we now discuss in detail.

     The resolution of these G270M observations is not sufficient to fully
resolve the LISM absorption.  This makes it difficult to identify the number
of LISM components, and in some cases also makes it difficult to separate the
LISM absorption from the intrinsic line profile.  Previously observed slowly
rotating K dwarfs have Mg~II profiles with self-reversals, i.e., the profiles
are double-peaked \citep{jll96,bew98}.
For HD~82558, HD~82443, and HD~17925 it is difficult to tell how much of
the apparent absorption is actually interstellar and how much of it is just
due to a self-reversal of the stellar line profile, although analyzing
the h and k lines together helps to solve this problem since the relative
depth of the possible self-reversals should be about the same for both lines,
while the depth of LISM absorption should be larger for the k line.

     We initially tried using Gaussians to model the Mg~II profiles of
HD~82443.  The location of the observed absorption feature is such that it is
clearly associated with the Local Interstellar Cloud (see below), but when
the intrinsic stellar line profiles are assumed to be Gaussian, the Doppler
parameter we measure for the LISM absorption is significantly larger than is
typically found for this cloud, and when the h and k lines are fitted
separately, the k line column density is somewhat smaller than that of the h
line.  Both of these properties suggest that Gaussians overestimate the
amount of flux overlying the LISM absorption.  Thus, we experimented with
profiles with self-reversals of various depths, which were created manually,
point-by-point.  In our best fit shown in Figure 2, the stellar profile has
only a very weak self-reversal.

     The Mg~II lines of HD~17925 proved to be the hardest to model.  The
analysis was similar to that described above for HD~82443, but the apparent
existence of two LISM components (see below) and the greater depth of the
absorption make the HD~17925 analysis more difficult and therefore more
uncertain.  Nonetheless, we believe that the Mg~II lines of HD~17925 do have
self-reversals.  However, the HD~17925 profile shown in Figure 2 has a very
strange property --- the red peak is stronger than the blue peak.  To the
best of our knowledge, all other self-reversed Mg~II line profiles observed
from cool main sequence stars have stronger {\em blue} peaks
\citep{bew96a,jll96,bew98}, as is also apparently the case for HD~82443.

     We cannot completely rule out the possibility that this is just a
manifestation of the difficulty in separating the LISM absorption and
self-reversal of HD~17925's Mg~II lines.  However, a more likely
explanation is that HD~17925 is actually a binary star, with Mg~II emission
from an unresolved secondary star being responsible for the stronger red peak.
Support for this assertion comes from observations of HD~17925's photospheric
absorption lines, which have variable widths that also suggest the presence
of a companion star \citep{gwh95,fcf97}.

     Rotational broadening is so severe for HD~197890 that the narrow
absorption seen near the center of its Mg~II lines is obviously entirely
interstellar, and for HD~1405 and HD~220140 the absorption is located far
enough from the center of the line that it is also clearly interstellar.
Thus, for these stars we can estimate the stellar line profile simply by
interpolating over the absorption, which we do by performing a polynomial
fit to data points on both sides of the absorption.

     The Mg~II lines of HD~1405 and HD~220140 clearly do not have
self-reversals.  This could be due in part to rotational smearing.
However, since these stars are very active it might also be a
stellar manifestation of a property of solar Mg~II lines observed from active
regions, where the depth of the self-reversals tends to be less than than
observed in quiet solar regions \citep{gad77}.
In any case, HD~82558 is rotating even faster than HD~1405 and HD~220140, and
is at least as active as these stars, so it is reasonable to presume that the
Mg~II lines of HD~82258 also do not possess self-reversals.  Thus, for
HD~82558 we use Gaussian fits to estimate the profiles of the Mg~II lines
(see Fig.\ 2).

\subsection{Fitting the Interstellar Mg~II Lines}

     Once the stellar Mg~II profiles have been estimated, we fit the LISM
absorption using standard techniques.  The atomic parameters needed for this
procedure are taken from \citet{dcm91}.  We use Voigt functions to represent
the opacity profiles of the fitted absorption.  Before being compared with
the data, the absorption profile is convolved with the line spread function
appropriate for SSA GHRS observations, which we assume to be a Gaussian with
a width of 3.7 pixels \citep{rlg94}.  In Figure 2, dotted lines show the
absorption components before convolution, and thick solid lines that fit the
data show the absorption after correcting for instrumental broadening.

     For each absorption component there are three parameters:  the central
velocity ($V_{\rm Mg}$), the column density ($N_{\rm Mg}$), and the Doppler
parameter ($b_{\rm Mg}$).  The Doppler parameter is related to the temperature
($T$) and nonthermal velocity ($\xi$) of the interstellar material by the
equation $b^{2} = 0.0165T/A + \xi^{2}$, where $b$ has units of
km~s$^{-1}$ and $A$ is the atomic weight of the element in question ($A=24$
for Mg).  Columns 3--5 of Table 3 list the parameters for the Mg~II fits.
Before deciding on these final fit parameters and their uncertainties, we
tried fitting the h \& k lines both independently and simultaneously to see
how the parameters change.  This gives us an idea of the magnitude of the
systematic errors involved in the fits, allowing us to more accurately
estimate the uncertainties in the various fit parameters.  Note that these
are just measurement uncertainties and do not include possible errors in the
wavelength calibrations (see \S 2).

     In the heliocentric rest frame, the Local Interstellar Cloud (LIC) in
which the Sun is embedded is flowing toward Galactic coordinates
$l=186^{\circ}$ and $b=-16^{\circ}$ at a velocity of about 25.7 km~s$^{-1}$
\citep{mw93,rl95}.  Using this vector we compute
the projected LIC velocities, V(LIC), for the lines of sight toward our K
dwarfs.  These velocities are listed in the second column of Table 4 and can
be compared with the measured $V_{\rm Mg}$ velocities in Table 3 to
identify which components are from LIC material.

     For most lines of sight we expect to detect absorption at the expected
LIC velocity.  However, the Sun is less than 0.2 pc from the edge of the LIC
in the general direction of the Galactic Center \citep{bew00}.
For lines of sight in this direction LIC absorption is often not observed,
but absorption from a different cloud called the ``G cloud'' is generally
detected.  The G cloud also has an accurately measured vector with a
direction similar to that of the LIC ($l=186^{\circ}$, $b=-16^{\circ}$), but
with a slightly faster speed of 29.4 km~s$^{-1}$ \citep{rl92}.
In column 3 of Table 4 we list the velocities, V(G), expected for G cloud
material for our studied lines of sight.

     For HD~220140 and HD~82443, the Mg~II lines can be fitted with only
one LISM absorption component, which has a velocity consistent with the LIC
(see Fig.\ 2 and Table 3).  Previous analyses of LIC and G cloud Mg~II
absorption lines observed with high spectral resolution have found that the
Doppler parameters are typically in the range $b_{\rm Mg}=2.2-3.5$ km~s$^{-1}$
\citep*{jll96,ard97,np97,bew98,bew00}, and the $b_{\rm Mg}$ values reported
for HD~220140 and HD~82443 in Table 3 fall nicely within this range.

     The Mg~II absorption seen toward the other four stars cannot be modeled
with only one component.  Thus, for these stars we model the absorption with
two components, although the limited resolution of our data means that even
more components could be present.  For HD~197890 the need for two
components is obvious since the components are well separated (see Fig.\ 2).
For HD~82558, HD~1405, and HD~17925, the evidence for the existence of two
components is more subtle.  When the absorption seen toward these stars is
fitted with only one component, the central velocity is consistent with
neither V(LIC) nor V(G), and the Doppler parameter is very large
($b_{\rm Mg}>5$ km~s$^{-1}$).  However, in the two component fits one of the
components does lie at the expected LIC velocity, and the Doppler parameters
are within the expected range.  Thus, we consider the two component models
presented in Figure 2 and Table 3 to be our best interpretations of the data.
However, because the components are so highly blended, it was necessary to
reduce the number of free parameters of these fits, so we have forced the
Doppler parameters of the two components to be the same.  This is indicated
in Table 3 by the use of parentheses around the $b_{\rm Mg}$ values of the
second component.

     The second column of Table 3 indicates which components are identified
with LIC absorption.  The velocity of HD~82558's Component 1 is consistent
with both the LIC and G clouds, but it is very difficult to associate it with
either for reasons that will become clear after the Lyman-$\alpha$ lines are
analyzed (see \S 3.3).  For the multicomponent fits, we have included a line
in Table 3 in which total column densities are listed.  Note that for the
highly blended components, the total Mg~II column density can be computed
more accurately than the column densities of the individual components.

     Only one of our stars lies anywhere near the Galactic Center direction:
HD~197890, with $l=6^{\circ}$ and $b=-38^{\circ}$.  For this line of sight we
would have {\em a priori} expected to see G cloud aborption but no LIC
absorption.  The V(LIC) and V(G) values in Table 4 are about the same
(${\rm V(LIC)}=-14.9$ km~s$^{-1}$ and ${\rm V(G)}=-15.2$ km~s$^{-1}$).
However, the $V_{\rm Mg}$ values listed in Table 3 for the two components
are $V_{\rm Mg}=-23\pm2$ km~s$^{-1}$ and $V_{\rm Mg}=0\pm1$ km~s$^{-1}$,
inconsistent with both the LIC and G clouds.  We cannot rule out the
possibility that Component 1 is actually an unresolved combination of two
components.  The $b_{\rm Mg}$ value for Component 1 is indeed suspiciously
high, although with a large uncertainty
($b_{\rm Mg}=4.4\pm 1.4$ km~s$^{-1}$).  However, the V(LIC) and V(G)
velocities lie at the very red edge of the observed Component 1 absorption
(see Fig.\ 2), meaning that the LIC and G clouds cannot account for a
significant amount of this absorption.

     In order to confirm this finding, we processed archival GHRS Mg~II
observations of AU~Mic, an M0 V star 9.94~pc from the Sun and only
$5^{\circ}$ from HD~197890.  The AU~Mic Mg~II h line spectrum is compared
with the HD~197890 data in Figure 3.  The AU~Mic spectrum was taken with the
Ech-B grating and therefore has a higher resolution than the G270M HD~197890
spectrum, but it was observed through the large science aperture (LSA) prior
to the installation of the COSTAR corrective optics into HST, meaning the
resolution is significantly degraded relative to post-COSTAR Ech-B data
\citep{srh95}.

     Only one LISM absorption component is seen toward AU~Mic, corresponding
to Component 1 seen toward HD~197890, although we had to arbitrarily shift
the AU~Mic spectrum by $-1$ km~s$^{-1}$ to make the absorption line up
perfectly.  Such a shift is within the combined uncertainties of the
wavelength calibrations of the two Mg~II spectra.  A difference in the
projected cloud velocity between the two stars might also
contribute to the slight velocity difference, even though the stars are
only $5^{\circ}$ apart.

     The fit to the HD~197890 Mg~II h line in Figure 2 is reproduced in
Figure 3, and we also compare the Component 1 absorption component
with the AU~Mic data, where we have used a line spread function appropriate
for pre-COSTAR LSA Ech-B data to correct for instrumental smoothing
\citep{rlg93}.  The reasonably good fit to the AU~Mic data
demonstrates that the Component 1 cloud does not extend beyond AU~Mic's
9.94~pc distance.  Otherwise we would see more absorption from the cloud
toward HD~197890, which lies 44.4~pc away.  The Component 2 cloud, on the
other hand, must lie beyond AU~Mic, since no Component 2 absorption is seen
in the AU~Mic spectrum.

     The dashed lines in Figure 3 show the expected location of G cloud
absorption, with the predicted LIC absorption being at about the same
velocity.  As mentioned above, the LIC and G clouds are clearly not
significant contributors to the Component 1 absorption.  Thus, neither the
LIC nor the G cloud is detectable toward HD~197890 or AU~Mic.  The
non-detection of the LIC is expected (see above), but the absence of G cloud
absorption is somewhat surprising.  Apparently the G cloud does not extend
very far to southern Galactic latitudes, explaining why it is not seen toward
either HD~197890 ($b=-38^{\circ}$) or AU~Mic ($b=-37^{\circ}$).  We estimate
an upper limit of $\log N_{\rm Mg}<12.0$ for the LIC and G clouds along this
line of sight.  Based on previous measurements of the Mg~II/H~I ratio in the
LIC and G clouds \citep*{ard97,np97,bew98,bew00},
this corresponds to upper limits for hydrogen of
$\log N_{\rm H}<17.5$ for the LIC and $\log N_{\rm H}<16.8$ for the G cloud.

\subsection{The Lyman-$\alpha$ Lines}

     The Lyman-$\alpha$ lines of the six K dwarfs are displayed in Figure 4.
The stellar emission lines are contaminated by broad interstellar H~I
absorption centered at a rest wavelength of 1215.670~\AA, and weaker
deuterium (D~I) absorption 0.33~\AA\ blueward of the H~I absorption.
The figure also shows our best estimates for the shapes of the
uncontaminated stellar Lyman-$\alpha$ lines, and our best fits to the LISM
absorption, which are discussed in detail below.  The H~I fit parameters are
listed in columns 6--8 of Table 3.

     The opacity of interstellar Lyman-$\alpha$ is so large that the H~I
absorption has wings that extend well beyond the saturated core of the
absorption.  When a value for the interstellar H~I column density
($N_{\rm H}$) is assumed, the shape of the wings of the intrinsic stellar
line profile outside the saturated core can be computed simply by
extrapolating upwards from the data, by multiplying the observed flux by
$\exp (\tau_{\lambda})$.  The shape of the center of the stellar line profile
can then be estimated by extrapolating between the wings, using the observed
shapes of the Mg~II lines as models for the appearance of Lyman-$\alpha$.
In this way, we create a set of model stellar Lyman-$\alpha$ profiles
designed to produce fits to the absorption for a set of assumed values
for $N_{\rm H}$.  We can then perform fits to the H~I and D~I absorption
lines using these profiles to see which ones actually yield good fits to the
data.

     We have used this procedure for analyzing Lyman-$\alpha$ lines
many times \citep*{jll96,np97,bew98,bew00}.  The analysis described in
\citet{np97} is particularly relevant, especially for the three
lines of sight discussed in that paper based on moderate resolution HST
observations of Lyman-$\alpha$ (31~Com, $\beta$~Cet, and $\beta$~Cas),
such as we have here.  The justification for the use of the Mg~II lines in
estimating the shape of the Lyman-$\alpha$ profiles is that Mg~II h \& k and
Lyman-$\alpha$ are all highly optically thick chromospheric lines that have
similar appearances in solar spectra.

     One of the main goals of the previous analyses mentioned above was
the measurement of the local D/H ratio.  However, it would be very difficult
to accurately measure D/H for our data because of the modest spectral
resolution, the relatively low signal-to-noise of the data, and the existence
of multiple LISM components for all but two of our stars.  Thus, in our
analysis we simply assume ${\rm D/H}=1.5\times 10^{-5}$, which is the
measured value for the LIC.  There is as yet
no conclusive evidence that D/H differs significantly from this value 
anywhere within the 50~pc distance range of our sample of stars \citep{jll98}.
The assumption of a D/H value means the observed D~I absorption constrains
the H~I column density as well as the D~I column density, which is very
helpful.

     As usual in Lyman-$\alpha$ analyses we assume H~I and D~I have the
same centroid velocities, and we assume the H~I and D~I Doppler parameters
are related by $b_{\rm D}=b_{\rm H}/2^{1/2}$.  This relation is applicable
if nonthermal velocities are negligible compared to the thermal velocities of
H~I and D~I, which previous work has demonstrated to to be a reasonably good
approximation for the warm LISM gas \citep*{jll96,np97,bew98,bew00}.
We take the fine structure of the H~I and D~I Lyman-$\alpha$ lines into
account in the analysis, although this has little effect on our results since
the two fine structure components are separated by only 1.3 km~s$^{-1}$.
With the assumptions mentioned above the D~I parameters are entirely
constrained by the H~I parameters, so the D~I parameters are not listed in
Table 3.

     In the case of multi-component fits, additional assumptions must be
made to limit the number of free fit parameters and ensure a unique fit to
the data.  For the two-component fits in Figure 4, we assume the velocity
separation of the components is the same as measured in the Mg~II fits.
We also assume the Doppler parameters of both components are identical, and
that the Mg~II/H~I abundance ratio is the same for both
components. This last assumption is potentially the most questionable since
Mg~II/H~I and Mg~II/D~I ratios are known to vary quite a bit in the LISM
\citep{np97}.  Thus, the column densities of the individual
components listed in Table 3 could be more uncertain than indicated by the
quoted error bars, but the {\em total} H~I column densities should not be
greatly sensitive to this assumption.  With the constraints listed
above, the parameters of the second component are completely constrained by
those of the first.  This interdependence is indicated in Table 3 by the use
of parentheses around the fit parameters of second components in two
component fits.

     With all the assumptions listed above, each fit in Figure 4 actually
has only three free parameters.  By experimenting with fits computed using
various stellar line profile models, we collect a set of acceptable fits to
the data that we use to define the fit parameters and uncertainties listed
in Table 3.  The measured H~I velocities, $V_{\rm H}$, are found to be
consistent with the Mg~II velocities, $V_{\rm Mg}$, considering the
quoted measurement uncertainties and the uncertainties in the wavelength
calibrations of the spectra (see \S 2).

     In many previous analyses of high resolution Lyman-$\alpha$ spectra,
the LISM H~I absorption has been found to be blended with absorption
from heliospheric material around the Sun and/or astrospheric material
surrounding the observed star \citep*{jll96,bew96b,ard97,bew98,bew00}.
The lower quality of our data makes it difficult to search for such
absorption, but the two spectra most likely to show evidence for this
additional absorption are the ones with the lowest H~I column densities:
HD~220140 and HD~82443.  Although we do not try to model these two data sets
with heliospheric or astrospheric absorption, we note that the Lyman-$\alpha$
fits for these two stars do have some undesirable properties that might
indicate the presence of this extra absorption.

     The fit to the D~I line of HD~82443 in Figure 4 is not particularly good
and suggests that H~I may actually be redshifted relative to D~I.  Models
suggest that heliospheric H~I absorption should be redshifted relative to the
LIC absorption for all lines of sight through the heliosphere
\citep{vbb98,hrm00}, so it is
possible that heliospheric H~I is contributing to the Lyman-$\alpha$
absorption on the red side of the line, thereby inducing the apparent
redshift of H~I relative to D~I.

     For HD~220140, the observed D~I absorption appears to be somewhat
narrower than the fit.  Also, the measured H~I Doppler parameter is very
high ($b_{\rm H}=14.2\pm 0.5$ km~s$^{-1}$), suggesting a very hot LIC
temperature of $T=12,200\pm 800$ K.  Previous measurements of the LIC
temperature based on much higher quality data suggest temperatures no higher
than about 9500 K, and more typically in the $7000-9000$~K range
\citep{jll95,ard97,np97,bew98}.  Both of these properties could in principle
be explained by the broadening of the H~I Lyman-$\alpha$ absorption line by
heliospheric and/or astrospheric absorption, which results in overestimates
of $b_{\rm H}$ and $b_{\rm D}$.

     For the two component fits, we list total H~I column densities in
Table 3 in addition to the individual component values, as is also done for
Mg~II.  The total $N_{\rm H}$ values observed for each line of sight will
prove useful in analyzing the EUVE observations of the stars (see \S 1).
In the last column of Table 3, the total Mg~II and H~I column densities are
used to derive logarithmic Mg depletion values for the six lines of sight,
where
$D({\rm Mg})\equiv \log (N_{\rm Mg}/N_{\rm H})-\log ({\rm Mg/H})_{\odot}$.
The solar Mg/H ratio is $3.9\times 10^{-5}$ \citep{ea89}.
We do not list $D({\rm Mg})$ values for the individual components of the
two-component fits since the Mg~II/H~I abundance ratio was assumed to be
the same for both components in the H~I fits.

     For the single component HD~220140 and HD~82443 lines of sight, which
both sample the LIC and only the LIC, our measured Mg depletions
($D({\rm Mg})=-0.84\pm0.14$ and $D({\rm Mg})=-0.89\pm0.18$, respectively) are
reasonably close to previous LIC measurements based on high resolution HST
spectra, which typically show $D({\rm Mg})\approx -1.1$
\citep{ard97,np97,bew98}.  For the HD~1405 and HD~17925
lines of sight, which have second components in addition to the LIC component,
$D({\rm Mg})=-0.54\pm0.32$.  We propose that the second components observed
for these two lines of sight have less Mg depletion than the LIC, resulting
in these lower depletion values.

     \citet{sr00} collected measurements of LIC H~I column
densities for many lines of sight and used them to construct a three
dimensional model of the LIC.  In the last column of Table 4 we list the
LIC column densities predicted by this model for our six lines of sight.
For HD~1405, HD~220140, HD~82443, and HD~17925 these predictions agree
surprisingly well with the measured LIC column densities listed in Table 3,
providing support for the accuracy of the model.  For
HD~197890 there is no detected LIC component, as discussed in \S 3.2, and
the very low predicted LIC column density in Table 4 is consistent with this
result.  For HD~82558, there is a component close to the expected LIC
velocity, but the measured column density ($\log N_{\rm H}=18.75\pm 0.15$) is
almost two orders of magnitude larger than the prediction
($\log N_{\rm H}{\rm (LIC)}=16.89$).

     While some inaccuracies in the \citet{sr00} model are to
be expected, the HD~82558 discrepancy is extreme.  If HD~82558's
Component 1 is in fact the LIC, it would suggest the existence of a small
dense knot of material within the LIC or a very extended, narrow finger of
material along the HD~82558 line of sight.  Another possibility is that
it is actually G cloud material.  The V(G) velocity listed for HD~82558
in Table 4 does in fact agree with the Component 1 $V_{\rm Mg}$ and
$V_{\rm H}$ velocities better than the V(LIC) velocity does.  Furthermore,
although the G cloud is believed to lie mainly toward the Galactic Center,
crude sketches of the cloud's location from \citet{rl92},
\citet{rl95}, and \citet*{bew00} suggest that the cloud
could extend to the $l=245^{\circ}$ longitude of HD~82558.

     Unfortunately, the $D({\rm Mg})=-1.44\pm 0.21$ value measured toward
HD~82558 is much closer to typical LIC values (see above) than to the
$D({\rm Mg})\approx -0.4$ values previously measured for the G cloud
\citep*{np97,bew00}.  If the HD~82558 Component 1 is
in fact the G cloud, it would suggest that Mg depletions must vary
greatly within the cloud.  Perhaps the most likely interpretation of
Component 1 is that it is neither the LIC nor the G cloud, but is an entirely
different cloud with the same projected velocity.  The two clouds seen toward
HD~82558 must be quite dense and compact to account for such high column
densities.

     The $\log N_{\rm H}=19.05\pm 0.15$ column density observed toward
HD~82558 implies an average H~I density along this line of sight of
$n_{\rm H}\approx 0.20$ cm$^{-3}$.  {\em We believe that this is the highest
average density ever observed for any line of sight through the nearby LISM,}
which strengthens the argument that the absorption toward HD~82558 is from
neither the LIC nor the G cloud.  The high column density is clearly evident
just by looking at HD~82558's Lyman-$\alpha$ line in Figure 4.  The H~I
absorption is so broad that it is blended with the D~I absorption.  A high
column density is also indicated by the absence of any detected emission in
the EUVE MW (170--380 \AA) spectrum of this star.

     By comparison, the column density observed
toward HD~82443 is quite small ($\log N_{\rm H}=17.70\pm 0.15$), despite the
fact that HD~82443 and HD~82558 are both 18 pc away and the two stars are
only $39^{\circ}$ apart in the sky.  Perhaps even more impressive is that
HD~82558 is also only $39^{\circ}$ from the ``interstellar tunnel'' toward
$\epsilon$~CMa ($l=240^{\circ}$, $b=-11^{\circ}$), for which ISM column
densities are lower than $\log N_{\rm H}=18.0$ even for lines of sight as
long as 200~pc \citep{cg95,jvv95}.  In the future,
it would be interesting to sample other lines of sight near HD~82558 to
determine the angular extent to which the high column densities persist.

\subsection{Chromospheric Line Fluxes}

     Now that stellar Mg~II and Lyman-$\alpha$ profiles have been estimated
for the purpose of analyzing the LISM absorption, we can also measure the
fluxes of those Mg~II and Lyman-$\alpha$ profiles.  We measure these fluxes
by direct integration, and the results are tabulated in Table 5.  As
mentioned in \S 2, many previous analyses of LISM Mg~II and Lyman-$\alpha$
absorption features have also utilized K dwarf spectra obtained by HST.
However, stellar line fluxes have generally not been reported in the
literature since these studies were focused on the properties of the LISM.

     Therefore, in addition to the line fluxes of the six stars analyzed here,
we take this opportunity to list in Table 5 the Mg~II and/or Lyman-$\alpha$
line fluxes for seven previously analyzed K0--K5 dwarf stars.  References for
the interstellar Mg~II/Lyman-$\alpha$ absorption analyses of these lines of
sight are provided in the last column of Table 5.  The inclusion of the
additional fluxes in Table 5 is potentially very useful for studies of the K
dwarf rotation-activity relation because it extends the sample to longer
rotation periods than our six star sample by itself (see Table 1).

     Estimated uncertainties for the Mg~II and Lyman-$\alpha$ fluxes
are listed in Table 5 in the form of percent errors.  In \S 3.3
we described how uncertainties in the Lyman-$\alpha$ LISM parameters were
derived by experimenting with different stellar profiles.  The uncertainties
in the stellar Lyman-$\alpha$ fluxes are also estimated in this process.
As Table 5 suggests, uncertainties of 20--25\% are typical for the
Lyman-$\alpha$ fluxes.  For HD~197890, the uncertainty is a bit lower (15\%)
because the very large width of this star's Lyman-$\alpha$ line makes it
easier to interpolate over the LISM absorption (see Fig.\ 4).  A lower 15\%
uncertainty is also quoted for HD~26965 because for this star more of the
stellar profile is uncontaminated by LISM absorption due to a large velocity
separation between the stellar emission and LISM absorption
\citep[see][]{bew98}.  For HD~82558, the Lyman-$\alpha$ flux uncertainty is
quite large (40\%) because the very high column density of this line of sight
absorbs a larger percentage of the star's total line emission (see Fig.\ 4).

     The uncertainty in the Mg~II fluxes is not generally dominated by the
uncertainty in the ISM correction, but by the uncertainties in the extent
of the emission line.  The Mg~II lines lie on top of a photospheric
continuum, the shape of which is unknown because of the overlying
chromospheric emission.  However, it is likely that the photospheric
continuum is close to zero underneath much of the chromospheric line because
of absorption from Mg~II ions in the photosphere and lower chromosphere
\citep{jll78,bew96a}.  Thus, in calculating the
chromospheric Mg~II fluxes we assume no underlying photospheric continuum.
Inaccuracies in this assumption and uncertainties in the wavelength range
used to compute the emission line flux are the major sources of systematic
error in the flux measurements.  For early K stars the photospheric flux
level around 2800~\AA\ is fairly low (see Fig.\ 1), but we believe it still
results in uncertainties of a few percent.  For a few stars, especially
HD~17925, the LISM correction further increases the estimated Mg~II flux
uncertainty (see Table 5).

\section{Summary}

     We have presented HST/GHRS observations of six K dwarf stars, and have
analyzed the interstellar absorption seen in the Lyman-$\alpha$ and Mg~II
h \& k lines of these stars.  Having measured the properties of the LISM
absorption components seen in these data and measured stellar Lyman-$\alpha$
and Mg~II fluxes corrected for this absorption, we summarize our primary
results as follows:
\begin{description}
\item[1.] One of our stars, HD~17925, has Mg~II lines that appear to have
  double-peaked Mg~II lines with stronger red peaks than blue peaks.  This is
  very unusual since the Mg~II lines of cool main sequence stars generally
  have stronger blue peaks.  Although we cannot rule out the possibility that
  this is due to difficulties in removing the LISM absorption, we think it is
  more likely that emission from a companion star is contributing to the
  lines.
\item[2.] For four of the six stars (HD~1405, HD~220140, HD~82443, and
  HD~17925), a LISM component is identified that not only has a velocity
  consistent with the projected velocity of the LIC, but also has an H~I
  column density in excellent agreement with the predictions of the
  \citet{sr00} model of the LIC.  This result provides support
  for the accuracy of the column densities predicted by this LIC model.
\item[3.] Neither LIC nor G cloud absorption is observed toward HD~197890.
  Based on the direction of this line of sight, the LIC nondetection was
  expected \citep[e.g.,][]{sr00}, but the G cloud nondetection is
  somewhat surprising, and indicates that the G cloud does not extend to
  Galactic latitudes as far south as HD~197890 ($b=-38^{\circ}$).  This
  result was confirmed using GHRS Mg~II observations of AU~Mic, a star only
  $5^{\circ}$ from HD~197890.
\item[4.] Two LISM components are detected toward HD~82558, and one has a
  velocity close to both the LIC and G cloud velocities.  However, it
  is very difficult to associate the component with either the LIC or G
  clouds, mostly because of its extremely high column density.  The total H~I
  column density toward HD~82558 is $\log N_{\rm H}=19.05\pm 0.15$.  The
  implied average density along this 18.3~pc line of sight,
  $n_{\rm H}\approx 0.2$ cm$^{-3}$, is the largest such density measured from
  HST data for any line of sight through the nearby LISM.  Interestingly
  enough, the star lies only $39^{\circ}$ from $\epsilon$~CMa, a direction
  where column densities are notoriously low --- $\log N_{\rm H}\leq 18.0$
  for lines of sight as long as 200~pc.
\item[5.] In addition to the six stars analyzed here, we also measure
  Lyman-$\alpha$ and/or Mg~II fluxes for seven other K stars based on
  previously published LISM absorption analyses.  The line fluxes for the
  stars in this expanded data set promise to be very useful for studying
  the chromospheric rotation-activity relation of K dwarf stars.
\end{description}

\acknowledgments

We would like to thank F.\ Fekel for providing us with new spectral type
measurements of HD~197890 and HD~82558, and we would like to thank the
referee, S.\ Federman, for many useful comments.  We also thank NASA for
support under grant S-56500-D to the University of Colorado and NIST.

\clearpage

\begin{deluxetable}{llcccccc}
\tabletypesize{\small}
\tablecaption{List of K Dwarfs}
\tablecolumns{8}
\tablewidth{0pt}
\tablehead{
  \colhead{Star} & \colhead{Alternate} & \colhead{Spectral} & \colhead{$d$} &
    \colhead{$l$} & \colhead{$b$} & \colhead{$P_{rot}$} &
    \colhead{References}\\
  \colhead{} & \colhead{Name} & \colhead{Type} & \colhead{(pc)} &
    \colhead{(deg)} & \colhead{(deg)} & \colhead{(days)} & \colhead{}}
\startdata
\sidehead{Primary Sample}
HD 197890&``Speedy Mic''&K2-3 V&44.4&   6 &$-38$& 0.38 & 1,2 \\
HD 82558  & LQ Hya    & K2 V & 18.3 & 245 &  28 & 1.60 & 3,4 \\
HD 1405   & PW And    & K2 V & 21.9 & 115 &$-31$& 1.75 & 1,5 \\
HD 220140 & V368 Cep  & K2 V & 19.7 & 119 &  17 & 2.77 & 6,7 \\
HD 82443  & DX Leo    & K0 V & 17.7 & 201 &  46 & 5.38 & 8,9 \\
HD 17925  & EP Eri    & K2 V & 10.4 & 192 &$-58$& 6.76 & 8,10 \\
\sidehead{Additional Stars}
HD 36705  & AB Dor    & K1 V & 14.9 & 275 &$-33$& 0.51 & 11,12 \\
HD 22049  &$\epsilon$ Eri&K2 V&3.22 & 196 &$-48$& 11.7 & 10,13 \\
HD 155886 & 36 Oph A  & K1 V & 5.46 & 358 &   7 & 20.7 & 10,13 \\
HD 209100 &$\epsilon$ Ind&K5 V&3.63 & 336 &$-48$&  22  & 13,14 \\
HD 201091 & 61 Cyg A  & K5 V & 3.48 &  82 & $-6$& 35.4 & 10,13 \\
HD 26965  & 40 Eri A  & K1 V & 5.04 & 201 &$-38$& 37.1 & 13,14 \\
HD 128621 &$\alpha$ Cen B&K0 V&1.35 & 316 & $-1$&  42  & 13,14 \\
\enddata
\tablerefs{(1) F.\ C.\ Fekel 2000, private communication. (2) Cutispoto
  et al.\ 1997. (3) Fekel et al.\ 1986. (4) Jetsu 1993. (5) Hooten \& Hall
  1990.  (6) Bianchi et al.\ 1991. (7) Mantegazza et al.\ 1992. (8) Henry
  et al.\ 1995. (9) Messina et al.\ 1999. (10) Donahue et al.\ 1996. (11)
  Wichmann et al.\ 1998. (12) Innis et al.\ 1988. (13) Gliese \& Jahreiss
  1991. (14) Saar \& Osten 1997.}
\end{deluxetable}

\begin{deluxetable}{lcccclrc}
\tablecaption{Summary of GHRS Observations}
\tablecolumns{8}
\tablewidth{0pt}
\tablehead{
  \colhead{Target} & \colhead{Grating} & \colhead{Aperture} &
    \colhead{Spectral} & \colhead{Exposure} & \colhead{Date} &
    \colhead{Start} & \colhead{WAVECAL?} \\
  \colhead{} & \colhead{} & \colhead{} & \colhead{Range} &
    \colhead{Time} & \colhead{} & \colhead{Time} & \colhead{} \\
  \colhead{} & \colhead{} & \colhead{} & \colhead{(\AA)} &
    \colhead{(s)} & \colhead{} & \colhead{(UT)} & \colhead{}}
\startdata
HD 220140 & G270M & SSA & 2775--2822 &  323 & 1995 Sept.\ 22 &  4:39 & No  \\
          & G160M & SSA & 1202--1240 & 1185 & 1995 Sept.\ 22 &  5:38 & No  \\
HD 1405   & G270M & SSA & 2775--2822 &  754 & 1995 Sept.\ 29 & 10:24 & No  \\
          & G160M & SSA & 1202--1240 & 1293 & 1995 Sept.\ 29 & 10:41 & No  \\
HD 197890 & G270M & SSA & 2775--2822 &  646 & 1995 Nov.\ 7   &  4:23 & Yes \\
          & G160M & SSA & 1202--1240 & 1077 & 1995 Nov.\ 7   &  5:29 & No  \\
HD 82443  & G270M & SSA & 2775--2822 &  233 & 1995 Nov.\ 9   & 13:05 & Yes \\
          & G160M & SSA & 1202--1240 & 1293 & 1995 Nov.\ 9   & 13:27 & No  \\
HD 17925  & G270M & SSA & 2775--2822 &  431 & 1996 Jan.\ 9   &  4:18 & Yes \\
          & G160M & SSA & 1202--1240 & 1077 & 1996 Jan.\ 9   &  4:31 & No  \\
HD 82558  & G270M & SSA & 2775--2822 &  323 & 1996 May 4     &  8:07 & Yes \\
          & G160M & SSA & 1202--1240 &  754 & 1996 May 4     &  8:19 & No  \\
\enddata
\end{deluxetable}

\begin{deluxetable}{lcccccccc}
\tabletypesize{\footnotesize}
\tablecaption{Measured ISM Properties\tablenotemark{a}}
\tablecolumns{9}
\tablewidth{0pt}
\tablehead{
  \colhead{Star} & \colhead{Component} & \colhead{$V_{\rm Mg}$} &
    \colhead{$b_{\rm Mg}$} & \colhead{$\log N_{\rm Mg}$} &
    \colhead{$V_{\rm H}$} & \colhead{$b_{\rm H}$} &
    \colhead{$\log N_{\rm H}$} & $D({\rm Mg})$ \\
  \colhead{} & \colhead{} & \colhead{(km s$^{-1}$)} &
    \colhead{(km s$^{-1}$)} & \colhead{} & \colhead{(km s$^{-1}$)} &
    \colhead{(km s$^{-1}$)} & \colhead{} & \colhead{}}
\startdata
HD 197890 &   1    & $-23\pm2$ & $4.4\pm1.4$ & $13.80\pm0.70$ &
                     $-27\pm3$ &$13.7\pm1.5$ & $18.30\pm0.15$ &   \nodata    \\
          &   2    &   $0\pm1$ & $3.6\pm1.5$ & $12.50\pm0.30$ &
                      $(-4)$   &   (13.7)    &     (17.00)    &   \nodata    \\
          & totals &  \nodata  &   \nodata   & $13.80\pm0.70$ &
                      \nodata  &   \nodata   & $18.30\pm0.15$ &$-0.09\pm0.72$\\
HD 82558  &   1    &   $6\pm2$ & $3.2\pm0.7$ & $12.90\pm0.25$ &
                       $7\pm2$ &$10.0\pm5.0$ & $18.75\pm0.15$ &   \nodata    \\
          &   2    &  $14\pm2$ &    (3.2)    & $12.90\pm0.25$ &
                      $(15)$   &   (10.0)    &     (18.75)    &   \nodata    \\
          & totals &  \nodata  &   \nodata   & $13.20\pm0.15$ &
                      \nodata  &   \nodata   & $19.05\pm0.15$ &$-1.44\pm0.21$\\
HD 1405   &  LIC   &  $10\pm2$ & $2.5\pm0.6$ & $13.10\pm0.50$ &
                       $7\pm2$ &$13.9\pm1.2$ & $18.05\pm0.10$ &   \nodata    \\
          &   2    &   $2\pm2$ &    (2.5)    & $13.10\pm0.50$ &
                      $(-1)$   &   (13.9)    &     (18.05)    &   \nodata    \\
          & totals &  \nodata  &   \nodata   & $13.40\pm0.30$ &
                      \nodata  &   \nodata   & $18.35\pm0.10$ &$-0.54\pm0.32$\\
HD 220140 &  LIC   &   $5\pm1$ & $2.7\pm0.3$ & $12.70\pm0.10$ &
                       $7\pm1$ &$14.2\pm0.5$ & $17.95\pm0.10$ &$-0.84\pm0.14$\\
HD 82443  &  LIC   &  $11\pm1$ & $3.0\pm1.0$ & $12.40\pm0.10$ &
                      $11\pm1$ &$12.0\pm0.5$ & $17.70\pm0.15$ &$-0.89\pm0.18$\\
HD 17925  &  LIC   &  $19\pm1$ & $3.4\pm1.0$ & $13.00\pm0.40$ &
                      $20\pm1$ &$11.8\pm0.8$ & $17.95\pm0.10$ &   \nodata    \\
          &   2    &   $9\pm2$ &    (3.4)    & $12.50\pm0.30$ &
                      $(10)$   &   (11.8)    &     (17.45)    &   \nodata    \\
          & totals &  \nodata  &   \nodata   & $13.10\pm0.30$ &
                      \nodata  &   \nodata   & $18.05\pm0.10$ &$-0.54\pm0.32$\\
\enddata
\tablenotetext{a}{Quantities in parentheses are not free parameters of the fit
  (see text).}
\end{deluxetable}

\begin{deluxetable}{lccc}
\tablecaption{Predicted ISM Properties}
\tablecolumns{4}
\tablewidth{0pt}
\tablehead{
  \colhead{Star} & \colhead{V(LIC)} & \colhead{V(G)} &
    \colhead{$\log N_{\rm H}{\rm (LIC)}$\tablenotemark{a}} \\
  \colhead{} & \colhead{(km s$^{-1}$)} & \colhead{(km s$^{-1}$)} & \colhead{}}
\startdata
HD 197890 & $-14.9$ & $-15.2$ & 16.53 \\
HD 82558  &   $7.9$ &   $7.2$ & 16.89 \\
HD 1405   &  $10.5$ &  $13.4$ & 18.02 \\
HD 220140 &   $6.9$ &   $7.7$ & 17.98 \\
HD 82443  &  $11.3$ &  $10.9$ & 17.59 \\
HD 17925  &  $19.1$ &  $23.1$ & 17.91 \\
\enddata
\tablenotetext{a}{Predicted column density from Redfield \& Linsky (2000).}
\end{deluxetable}

\begin{deluxetable}{lccccccc}
\tabletypesize{\footnotesize}
\tablecaption{Chromospheric Line Fluxes}
\tablecolumns{8}
\tablewidth{0pt}
\tablehead{
  \colhead{Star} &
    \multicolumn{3}{c}{Fluxes ($10^{-12}$ ergs cm$^{-2}$ s$^{-1}$)} & &
    \multicolumn{2}{c}{\% Errors} & \colhead{References}\\
  \cline{2-4} \cline{6-7}\\
  \colhead{} & \colhead{Mg II k} & \colhead{Mg II h} &
    \colhead{Lyman-$\alpha$} & & \colhead{Mg II} & \colhead{Lyman-$\alpha$} &
    \colhead{}}
\startdata
\sidehead{Primary Sample}
HD 197890 & 0.91 & 0.73 & 1.85 & & 5 & 15 & 1 \\
HD 82558  & 2.83 & 2.27 & 4.01 & & 5 & 40 & 1 \\
HD 1405   & 1.42 & 1.12 & 2.31 & & 3 & 25 & 1 \\
HD 220140 & 3.07 & 2.45 & 2.99 & & 3 & 25 & 1 \\
HD 82443  & 3.29 & 2.58 & 2.31 & & 4 & 20 & 1 \\
HD 17925  & 7.57 & 5.83 & 6.05 & & 8 & 25 & 1 \\
\sidehead{Additional Stars}
HD 36705  & 5.97 & 4.79 &\nodata& &5 &\nodata& 2 \\
HD 22049  &\nodata&\nodata&48.8& &\nodata&25& 3 \\
HD 155886 & 6.39 & 4.60 & 14.2 & & 4 & 20 & 4 \\
HD 209100 &\nodata&\nodata&31.0& &\nodata&20& 5 \\
HD 201091 & 8.32 & 5.84 & 17.2 & & 3 & 20 & 6 \\
HD 26965  & 6.71 & 4.91 & 6.93 & & 4 & 15 & 6 \\
HD 128621 & 164  & 121  & 150  & & 4 & 25 & 7 \\
\enddata
\tablerefs{(1) This paper. (2) Brandt et al.\ 2000. (3) Dring et al.\ 1997.
  (4) Wood et al.\ 2000. (5) Wood et al.\ 1996b. (6) Wood \& Linsky 1998.
  (7) Linsky \& Wood 1996.}
\end{deluxetable}

\clearpage

\begin{figure}
\plotfiddle{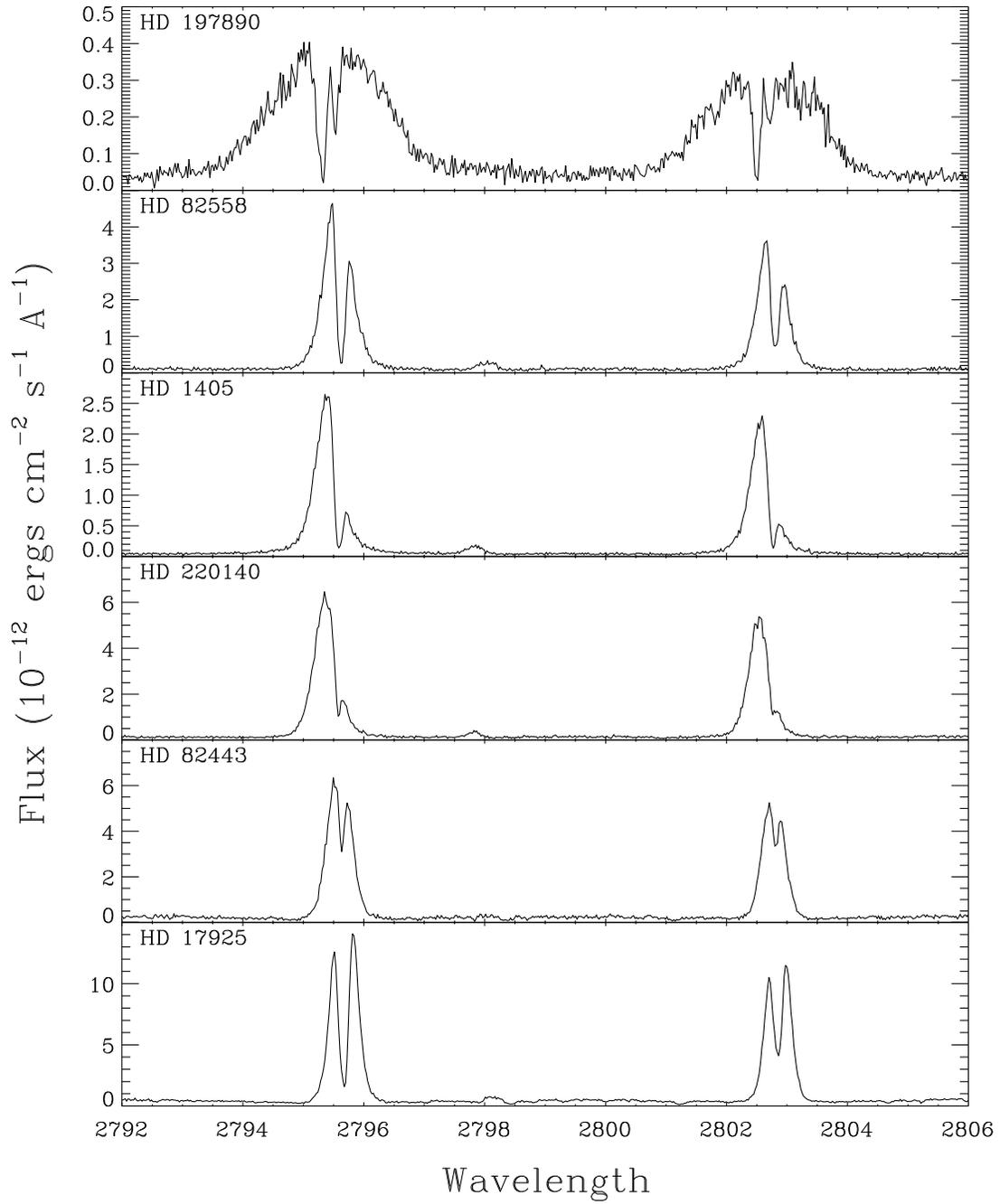}{6.5in}{0}{80}{80}{-340}{10}
\caption{The Mg~II h \& k lines of the six K dwarfs, with rest wavelengths
  in air of 2802.705~\AA\ and 2795.528~\AA, respectively.}
\end{figure}

\clearpage

\begin{figure}
\plotfiddle{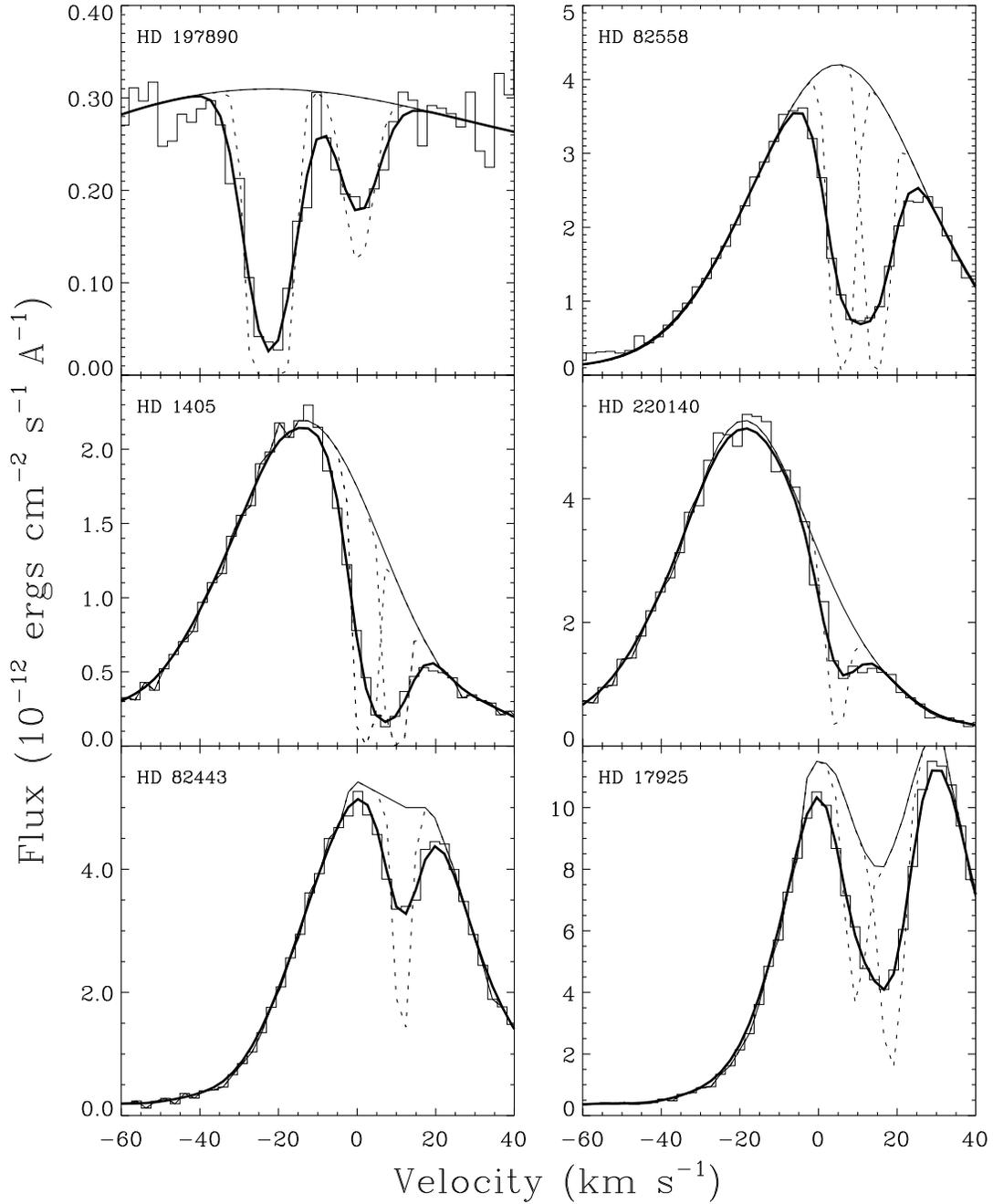}{6.5in}{0}{80}{80}{-340}{10}
\caption{Fits to the interstellar absorption observed in the Mg~II h lines
  of the six K dwarf stars.  For each star, the assumed stellar emission
  profile is a thin solid line, the fitted absorption line or lines are
  dotted lines, and the convolution of these absorption lines with the
  instrumental profile is the thick solid line, which fits the data.}
\end{figure}

\clearpage

\begin{figure}
\plotfiddle{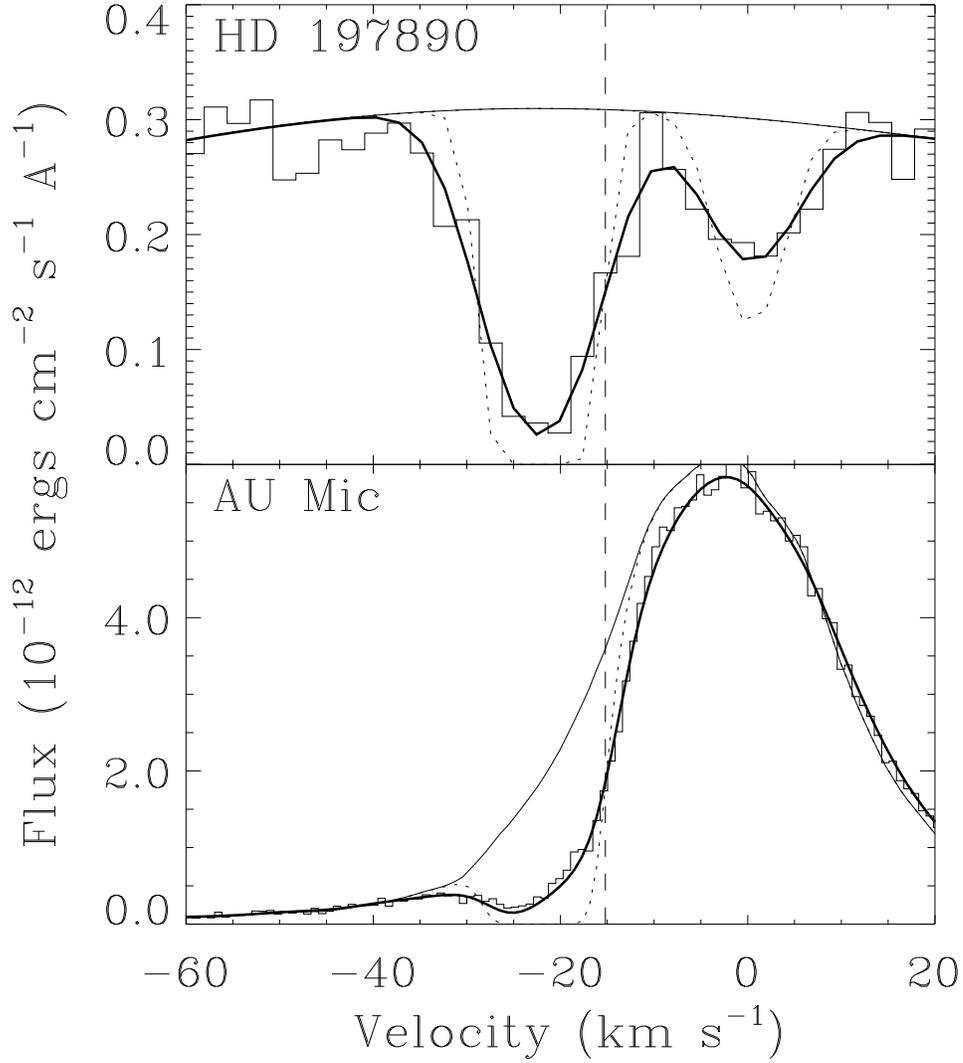}{4.5in}{0}{80}{80}{-270}{0}
\caption{The upper panel is a reproduction of the Mg~II h line fit to two
  absorption components of HD~197890 from Fig.\ 2, and in the bottom panel
  the fitted absorption profile for the blue component is applied to
  observations of the Mg~II h line of AU~Mic, which is only $5^{\circ}$ from
  HD~197890.  The same absorption profile fits the absorption of both stars
  quite well.  The red absorption component  seen toward HD~197890 is not
  seen at all toward the closer star AU~Mic.}
\end{figure}

\clearpage

\begin{figure}
\plotfiddle{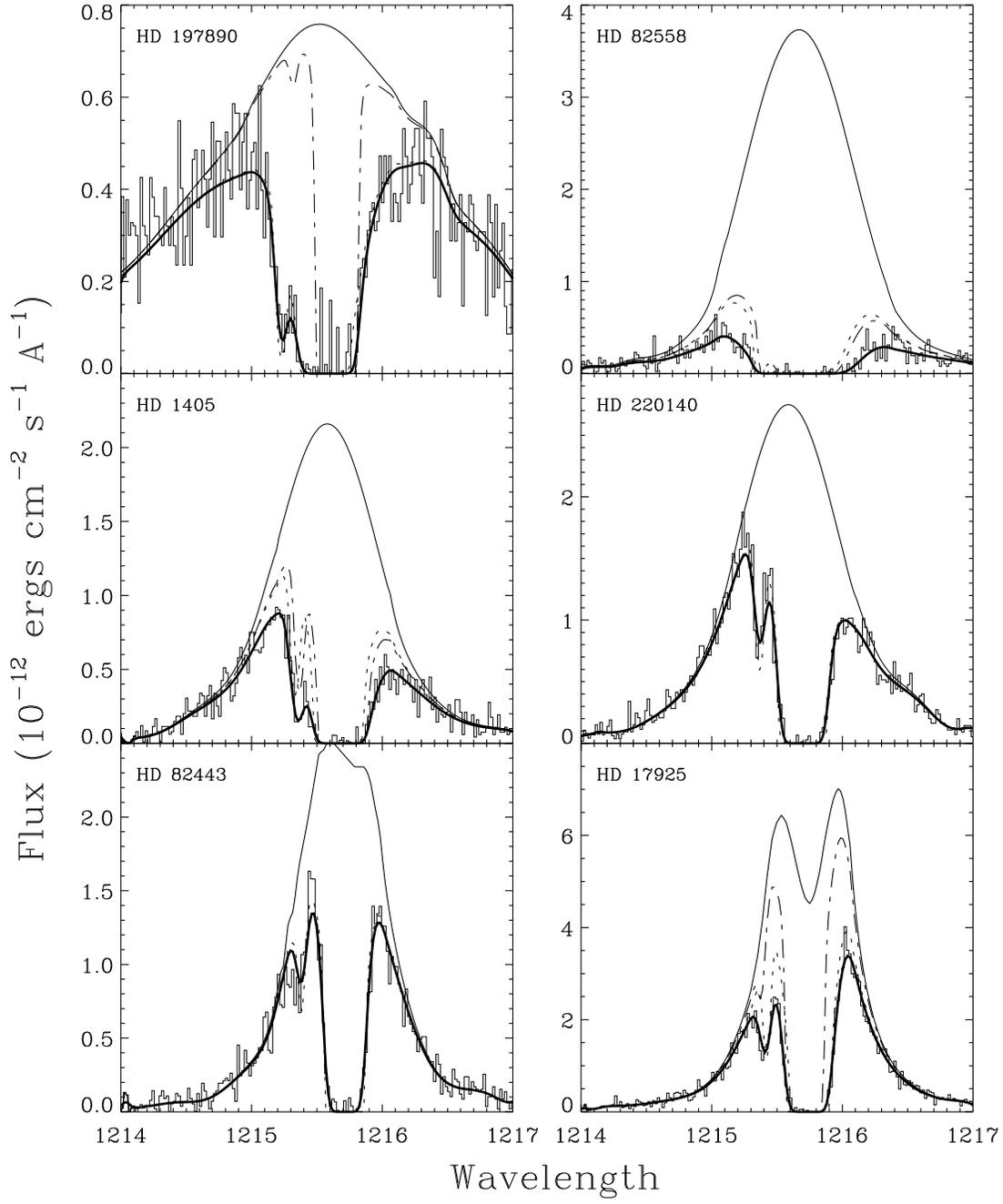}{6.5in}{0}{80}{80}{-340}{10}
\caption{Fits to the interstellar hydrogen (H~I) and deuterium (D~I)
  absorption observed in the Lyman-$\alpha$ lines of the six K dwarf stars,
  where the D~I absorption is the narrow absorption seen blueward of the
  broad, optically thick H~I absorption.  For each star, the assumed
  stellar emission profile is a thin solid line, the first fitted absorption
  component is a dotted line, the second absorption component (if there is
  one) is a dot-dashed line, and the convolution of these absorption
  components with the instrumental profile is the thick solid line, which 
  fits the data.}
\end{figure}

\end{document}